\documentclass[
reprint,
 amsmath,amssymb,
 aps,
floatfix,
]{revtex4-1}

\usepackage{ulem}
\usepackage{graphicx}
\usepackage{dcolumn}
\usepackage{bm}
\usepackage{subfigure} 
\usepackage{color}
\usepackage{soul}
\definecolor{BrickRed}{cmyk}{0, .89, .94, .28}
\definecolor{MyDarkBlue}{rgb}{0,0.08,0.45}
\definecolor{Mygray}{gray}{0.60}

\def\vac#1{\mathbf{#1}}
\def\be{\begin{equation}}
\def\ee{\end{equation}}
\def\beq{\begin{eqnarray}}
\def\eeq{\end{eqnarray}}

\begin{document}

\preprint{APS/123-QED}

\title{Optical concentrator from a hyperbolic liquid crystal metamaterial}

\author{Frankbelson dos S. Azevedo}
\email{dossanto2@univ-lorraine.fr}
\affiliation{Dynamique et Sym\'etrie, and Coll\`ege Doctoral ${\mathbb{L}}^4$ for Physics of Complex Systems, Laboratoire de Physique et Chimie Th\'eoriques, UMR Universit\'e de Lorraine - CNRS 7019,
 54506 Vand\oe uvre les Nancy, France}

\author{David Figueiredo}
\email{david.odf85@gmail.com}
\affiliation{ Departamento de F\'{\i}sica, CCEN, Universidade Federal da Para\'{\i}ba, 58051-900, Jo\~ao Pessoa, PB, Brazil}

\author{Fernando Moraes}
\email{fernando.jsmoraes@ufrpe.br}
\affiliation{Departamento de F\'{\i}sica, Universidade Federal Rural de Pernambuco, 
52171-900, Recife, PE, Brazil}

\author{Bertrand Berche}
\email{bertrand.berche@univ-lorraine.fr}
\affiliation{Dynamique et Sym\'etrie, Laboratoire de Physique et Chimie Th\'eoriques,  UMR Universit\'e de Lorraine - CNRS 7019,
 54506 Vand\oe uvre les Nancy, France}
\author{S\'ebastien Fumeron}
\email{sebastien.fumeron@univ-lorraine.fr}
\affiliation{Dynamique et Sym\'etrie, Laboratoire de Physique et Chimie Th\'eoriques,  UMR Universit\'e de Lorraine - CNRS 7019,
 54506 Vand\oe uvre les Nancy, France}

\date{\today}

\begin{abstract}
We examine the optical properties of two different configurations of a cylindrical device made from a hyperbolic metamaterial {with optical axis defined by circular and radial director fields.} {The hyperbolic metamaterial is an uniaxial anisotropic} medium for which the ratio between ordinary and extraordinary permittivities is negative, {leading to a particular effective geometry with two timelike coordinates in the metric (Kleinian signature)}. {By using differential geometry tools we are able to perform a} comparison between a simple geometrical optics treatment and {the} wave formalism {that} shows the concentration of light along the cylinder axis {for the case of the circular field configuration}, whatever 
{the} injection {conditions} are.
\end{abstract}

%\pacs{Valid PACS appear here}
                             
\maketitle

{Concentration of light has become a major issue in large number of applications ranging from solar energy harvesting to optical sensing. Different ways have been explored in order to focus light in the most efficient way. Historically, the first attempts were made by using lenses of dielectric matter, but {for a long time} diffraction phenomena forbid beam sizes below half of a wavelength. Plasmonic structures \cite{Memarzadeh2011} and optical metamaterials \cite{Pendry2000} are probably the best candidates to overcome the usual physical limits. These latter are artificial media {that} can be used in superlenses working beyond the diffraction limit. In this letter, we examine the possibility to focus light with a cylindric device made of a nematic liquid crystal which optically behaves as an anisotropic metamaterial. Performances of the device are discussed both in the geometrical optics limit and in the framework of wave theory.}

In {uniaxial} anisotropic media, light propagation is described in terms of two indices, $n_o$ the ordinary index and $n_e$, the extraordinary index. {For instance,} nematic liquid crystals are  {made of cigar-shaped molecules arranged in such a way that could exhibit properties of} uniaxial anisotropic media. At high temperatures, entropy maximization is realized by an isotropic (liquid) phase where all molecule orientations are equally likely, but when the temperature decreases, several phase transitions may occur. Among them, the appearance of a nematic phase with an orientational order is characterized by a specific direction measured by the director, $\vac n$, (a unit vector) along which  the molecules are aligned on average~\cite{RevModPhys.46.617}. The corresponding phase transition may be of second order or first order, depending in particular on space dimension and intermolecular interactions, and may even be, in low dimension, a topological phase transition governed by the presence of topological defects \cite{PhysRevE.72.031711}. 
In which concerns light propagation in the nematic phase, one has to distinguish between an ordinary ray, which propagates in such a way that the electric field of the electromagnetic wave  remains perpendicular to the director $\vac n$, and an extraordinary ray, for which $\vac E$ has a non vanishing component along $\vac n$. In the latter,  the Poynting vector $\vac S=\vac E\times\vac H$ differs in direction from that of the wave vector $\vac k$. In this case, energy velocity is thus governed by a {\bf ray index}
\be
N_r^2=n_o^2\cos^2\beta+n_e^2\sin^2\beta\label{Eq-RayIndex}
\ee
with $\beta=(\widehat{\vac n,\vac S})$ and differs from the phase velocity, governed by another index, the {\bf phase index}
\be
N_p^2=\frac{n_o^2n_e^2}{n_e^2\cos^2\gamma+n_o^2\sin^2\gamma}
\ee
with $\gamma=(\widehat{\vac n,\vac k})$ \cite{maxborn_optics}.
Fermat's principle states that the energy propagation for the extraordinary ray obeys the variational prescription
\be
\delta\int N_r(\vac r) dl=0, \label{Eq-FermatNematic}
\ee
where $dl$ is the element of arclength along the light path.

{We now analyze the light path in a hyperbolic liquid crystal metamaterial (HLCM) by Fermat's principle.} From the optical point of view, this is a uniaxial medium and the ordinary and extraordinary indices are given in terms of the permittivities, $n_o^2=\varepsilon_\perp$ and $n_e^2=\varepsilon_{||}$.  The material is made of an ordinary nematic liquid crystal which includes an admixture of metallic nanorods \cite{xiang}, which align along the director field $\vac n$ of the nematic phase, resulting in a negative permittivity $\varepsilon_{||}<0$ along that direction. {As shown in \cite{ward2005physical}, a model of negative index material cannot, strictly speaking, circumvent dispersion phenomena and losses. {The main consequence is that a concentrator designed from metamaterials should support evanescent waves and therefore, in order to be efficient, the length of the device has to be short enough to avoid that the losses cancel the field.} However, losses due to metallic plasmonic components might be offset by using gain media as pointed in \cite{bergman2003,noginov2008} (highly doped oxides with lower dissipation levels have also been considered \cite{naik2010, west2010}). {The low-loss limit for metamaterials has also been recently reached by working within the terahertz waveband \cite{suzuki2018}.} Hence, as customary in transformational optics (for example in the design of hyperlenses or invisibility cloaks \cite{veselago2002,leonhardt2009,kildishev2011}), we will consider metamaterials in homogenization regime for which losses have been reduced so that one may focus on kinematic aspects of light propagation.} 

\begin{figure}[h]
  \centering
(a)\includegraphics[width=.44\columnwidth]{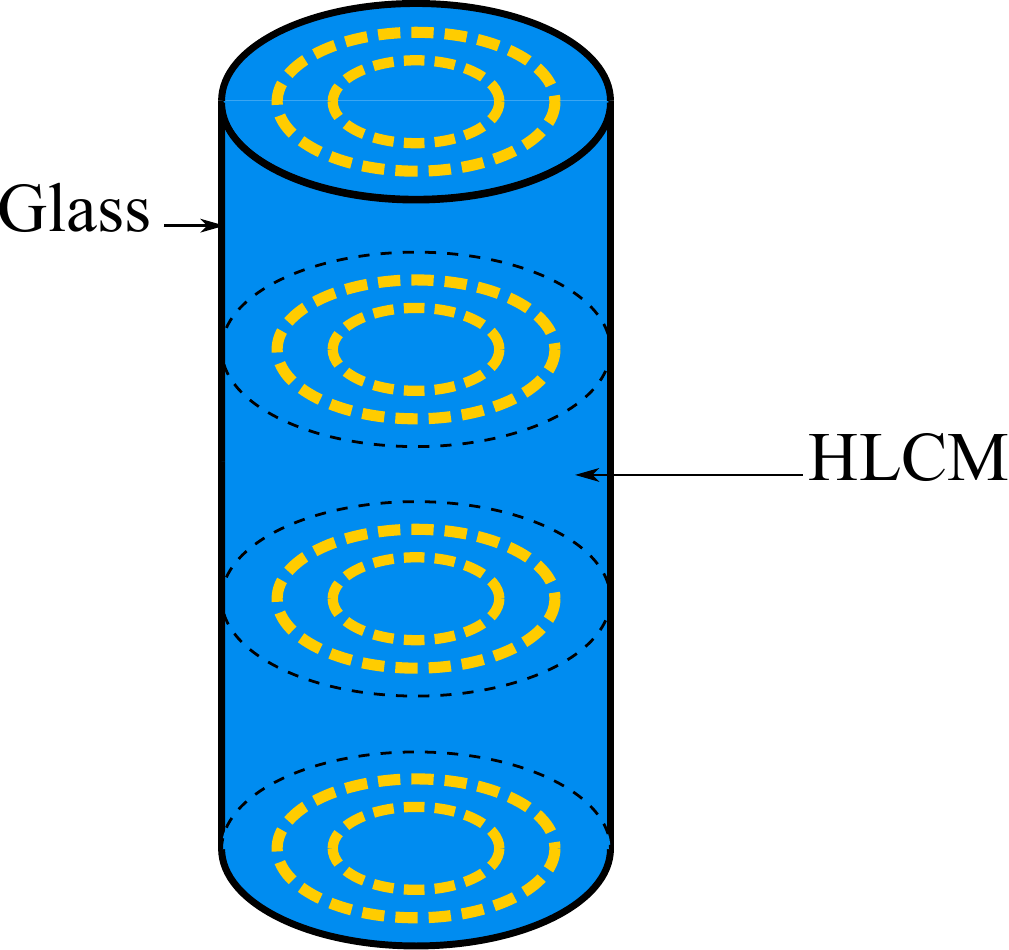}
  \hfill
(b)\includegraphics[width=.44\columnwidth]{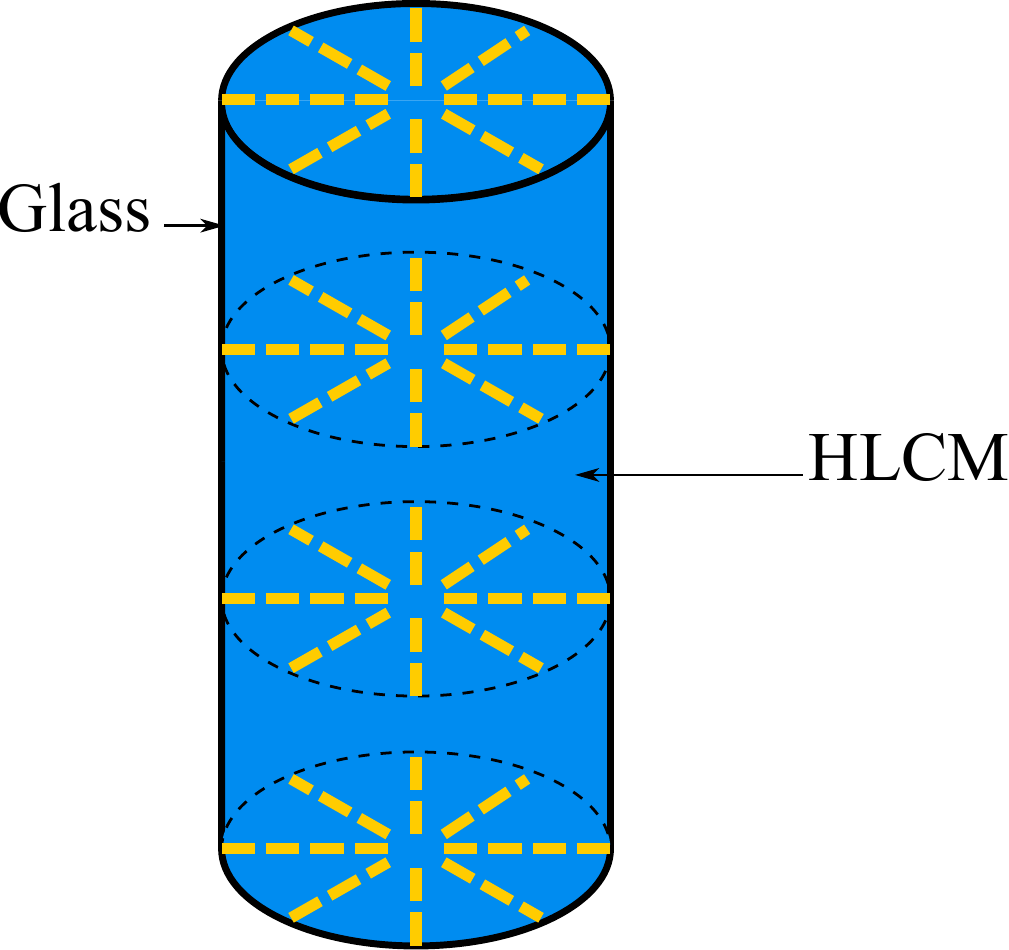}
  \caption{Two different cylindrical configurations of HLCM (according to director field arrangement) covered with a reflective material.   (a) Director field configuration for HLCM with optical axis as ${\hat{\bm n}}={\hat{\bm \phi}}$ (circular director field). 
  (b) Director configurations for HLCM with optical axis as ${\hat{\bm n}}={\hat{\bm r}}$ (radial director field).
  }
\label{fig1}
\end{figure}

We study two different types of HLCM configurations. First we consider the director field as ${\hat{\vac n}}={\hat{\bm \phi}}$ (circular director field, a hat is used to denote unit vectors), and then we consider the configuration ${\hat{\vac n}}={\hat{\bm r}}$  (radial director field), see {fig. \ref{fig1}}. 

Since the ray path is parametrized by the arc length $l$, the tangent unit vector along the path is thus $\vac t=\frac{d\vac r}{dl}$, and the angle $\beta$ which measures the tangent orientation w.r.t. the director is given by 
\be\cos\beta=\vac t\cdot\vac n.\label{Eq-CosBeta}\ee In cylindrical coordinates, ${\vac r}(l)=r \hat{\bm r} +z\hat{\vac z}$, thus $\vac t=\dot r\hat{\bm r} +r\dot\phi \hat{\bm \phi}+\dot z\hat{\vac z}$ where $\dot x=\frac{dx}{dl}$, and the normalization constraint follows $|\vac t|^2=\dot r^2+r^2\dot\phi^2+\dot z^2=1$. In the configuration (a), we thus have $\cos\beta=r\dot\phi$ from (\ref{Eq-CosBeta}), and $\sin^2\beta=\dot r^2+\dot z^2$ from the normalization constraint. The ray index (\ref{Eq-RayIndex}) in configuration (a) is then
\be
N_r^2=\varepsilon_\perp r^2\dot\phi^2-|\varepsilon_{||}|(\dot r^2+\dot z^2).\label{Eq-RayIndexConfigA}
\ee
In a similar manner, the ray index in configuration (b) is given by
\be
N_r^2=\varepsilon_\perp\dot r^2-|\varepsilon_{||}|(r^2\dot\phi^2+\dot z^2).\label{Eq-RayIndexConfigB}
\ee

Rescaling the coordinates according to $\rho=r \sqrt{|\epsilon_{\parallel}|}$, $\zeta=z \sqrt{|\epsilon_{\parallel}|}$, the elementary optical path $ds^2=N_r^2 dl^2$ writes as:
\be
ds^2=N_r^2 dl^2=-d\rho^2+\alpha^2\rho^2d\phi^2-d\zeta^2 ,\label{intConfigA}
\ee
for configuration (a) and 
\be
ds^2=N_r^2 dl^2=\alpha^2d\rho^2-\rho^2d\phi^2-d\zeta^2 ,\label{intConfigB}
\ee
for configuration (b), where $\alpha^2=\frac{\varepsilon_\perp}{|\varepsilon_{||}|}$. In both cases, the optical path can be elegantly reinterpreted in terms of non-Euclidean geometry: following a pioneering idea by Gordon \cite{Gordon1923analog}, light propagation inside a refractive medium occurs in a similar fashion to light propagation on a Riemannian manifold, the metric tensor of which is obtained from \cite{klinekay}:
\be
ds^2=g_{ij}dx^i dx^j
\ee
(here we used Einstein's summation convention on repeated indices). As will be illustrated in {what follows}, differential geometry is extremely powerful when dealing with calculations of light trajectories or with how to generalize the wave equation.

We now consider the geometrical optics limit and  start analyzing the path followed by light in the HLCM device with a purely circular director field (see {fig.} \ref{fig1}(a)), which is obtained from planar anchoring of the molecules at the boundaries. From the optical path (\ref{intConfigA}), the metric tensor is given in that case by:
\begin{eqnarray}
g_{ij}&=&
\begin{pmatrix}
-1&0&0\\
0&\alpha^2\rho^2&0\\
0&0&-1
\end{pmatrix}. \label{metric-a} 
\end{eqnarray}
Constants of motion are given by the Killing vectors, which correspond the cyclic variables of the metric. A quick examination of (\ref{metric-a}) reveals two Killing vectors, $(\partial_{\phi})^i=(0,1,0)^{\rm T}$ and $(\partial_{\zeta})^i=(0,0,1)^{\rm T}$ {(here $^{\rm T}$ denotes the transposition operation used to represent column vectors)}, associated with the covectors:
\begin{eqnarray}
&&(\partial_{\phi})_i=g_{ij}(\partial_{\phi})^j=(0,\alpha^2\rho^2,0), \\
&&(\partial_{\zeta})_i=g_{ij}(\partial_{\zeta})^j=(0,0,-1). 
\end{eqnarray}
These vectors obey the Killing equations, which give the constants of motion: 
\begin{eqnarray}
&&(\partial_{\phi})_i\frac{dx^i}{d\lambda}=\widetilde{C}, \\
&&(\partial_{\zeta})_i\frac{dx^i}{d\lambda}=A.
\end{eqnarray}
Here $\lambda$ is an affine parameter such that $ds^2=Bd\lambda^2$ with $B>0$ in order to preserve the causal regions in terms of the two variables. 
Denoting by $x'=\frac{dx}{d\lambda}$, one gets
\begin{eqnarray}
&&\zeta'=A, \label{pi1} \\
&&\rho^2\phi'=\frac{\widetilde{C}}{\alpha^2}=C. \label{pi2} 
\end{eqnarray}
In the remainder, propagation along increasing values of $z$ will be considered, so that $A>0$. A third constant of motion is obtained from energy conservation. The line element gives:
\begin{equation}
-(\rho')^2+\alpha^2\rho^2(\phi')^2-(\zeta')^2=B.\label{Eq17}
\end{equation}
{Note that Eq.~(\ref{Eq17}) can also be deduced from the eikonal equation $k_\mu k^\mu=0$.}
Substituting (\ref{pi1})-(\ref{pi2}) gives 

\begin{equation}
\frac{\rho'^2}{2}-\frac{\alpha^2 C^2}{2\rho^2}=\frac{-(A^2+B)}{2}=E. \label{energy-cons}
\end{equation}

It must be remarked that the particular form of that relation requires that the energy parameter $E\leq 0$. The resolution of that equation leads to:
\begin{equation}
\frac{\rho d\rho}{\sqrt{\rho_M^2-\rho^2}}=\pm K d\zeta, \label{eqofparam}
\end{equation}
where $\rho_M=\frac{\alpha |C|}{\sqrt{2|E|}}$ and $K=\frac{\sqrt{2|E|}}{A}>0$. Denoting by $\rho_0$ the radius at which the ray is injected ($\zeta=0)$, one finds the two families of solutions. The first family corresponds to the positive sign, which corresponds to rays of increasing radius:
\begin{equation}
\rho(\zeta)=\sqrt{\rho_M^2-((\rho_M^2-\rho_0^2)^{1/2}-K\zeta)^2}. \label{solution-typeI}
\end{equation}

Hence, orbits are confined with maximum radius $\rho_M$. We also see that a given ray reaches  $\rho_M$ for $z_M=\frac{(\rho_M^2-\rho_0^2)^{1/2}}{\sqrt{|\epsilon_{\parallel}|}K}$. For moderate injection angles, the light ray is not able to reach the outer metallic layer before falling down onto the defect core: in other words, rays undergo a total internal reflection for any frequency preserving the hyperbolic feature of the nematic phase. For large injection angles, light rays may  reach the outer metallic layer but will be reflected back inside the HLCM to finally converge on the core.

\begin{figure}[h]
\includegraphics[scale=0.5]{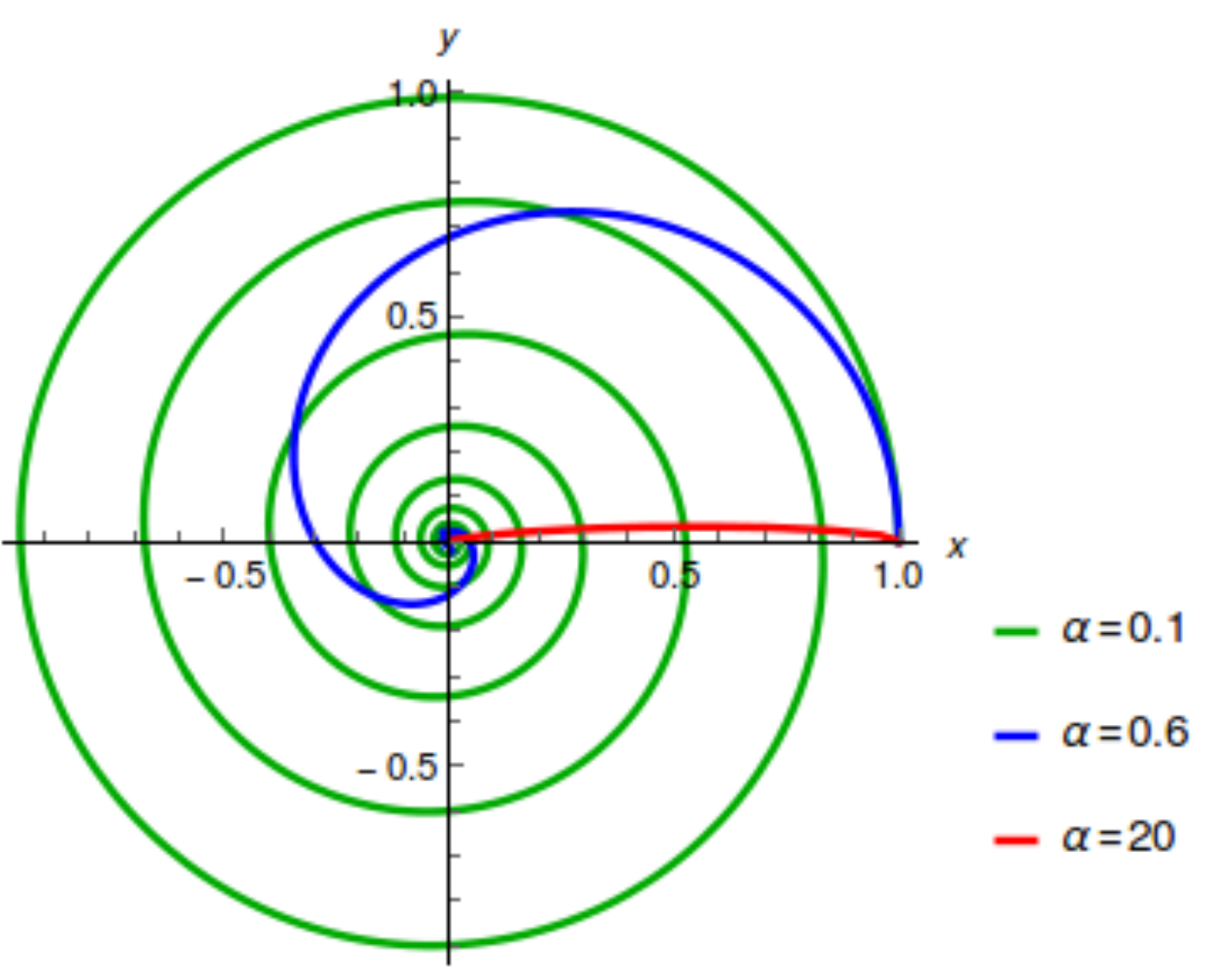} 
 \centering
\caption{Projection of the trajectories onto the $x-y$ plane with $\rho$ in units of $\rho_M$, for a few values of $\alpha$.} \label{plotsr0}
\end{figure}

\begin{figure}[h]
\centering
(a)\includegraphics[scale=0.29]{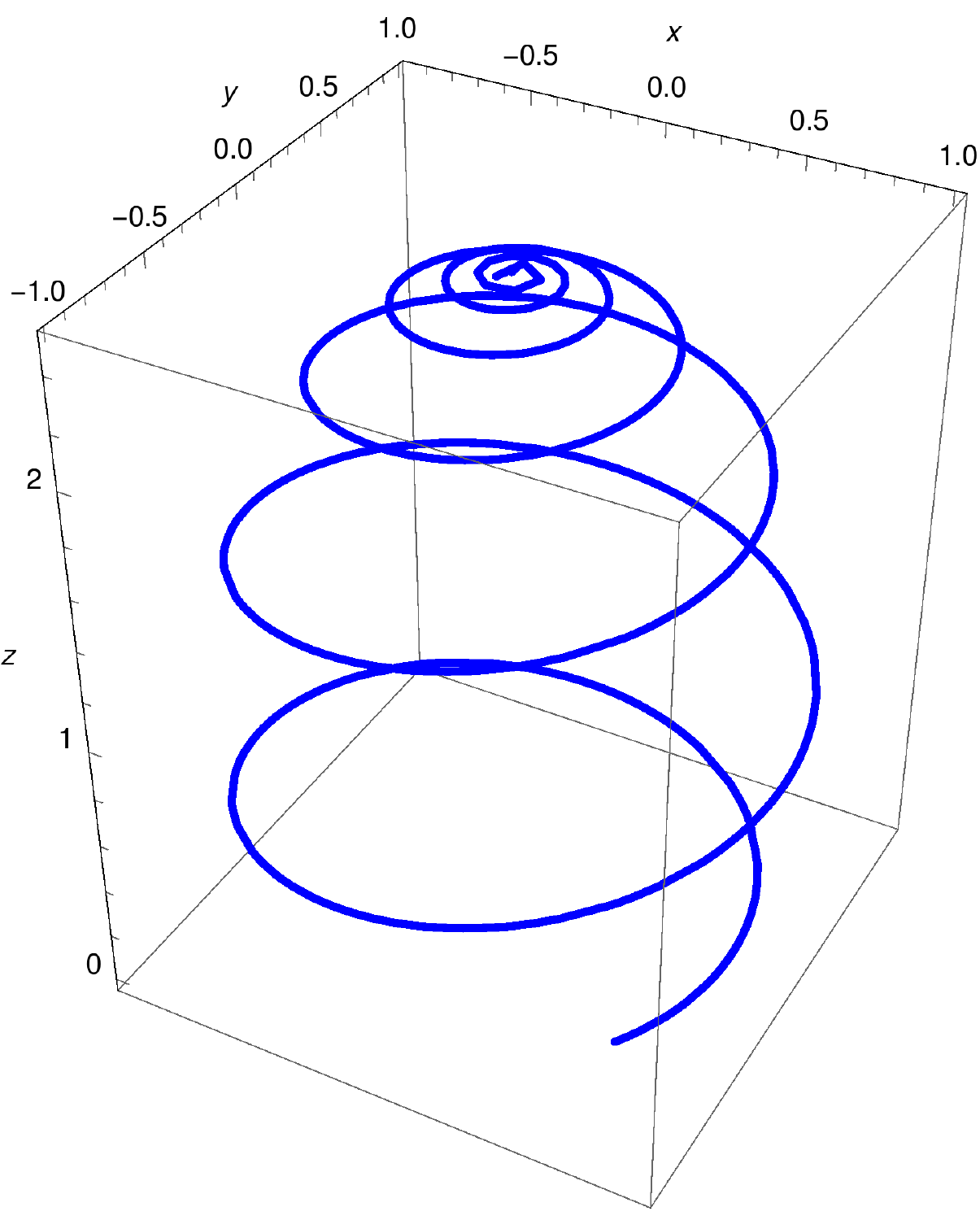}
\includegraphics[scale=0.29]{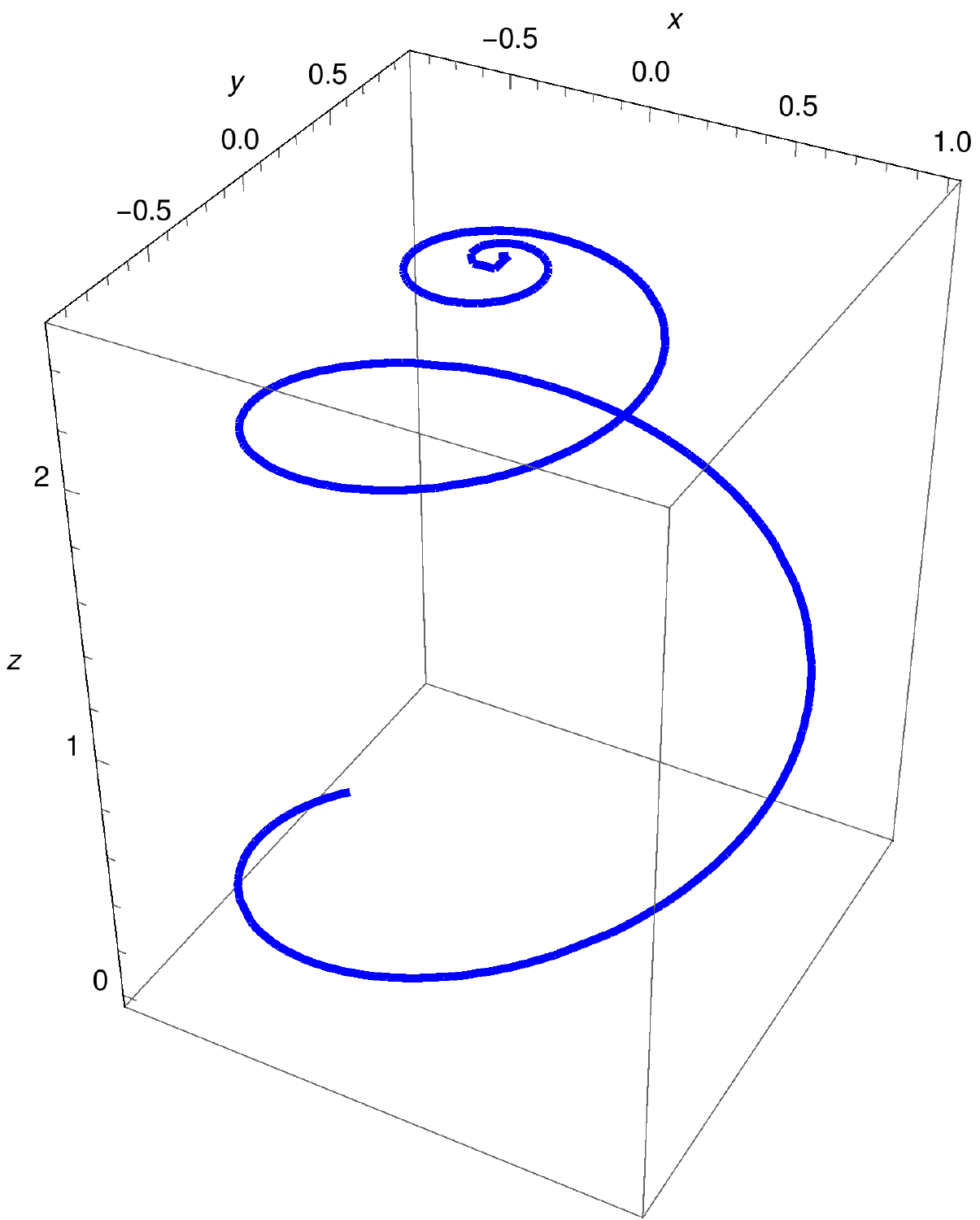} \\
(b)\includegraphics[scale=0.29]{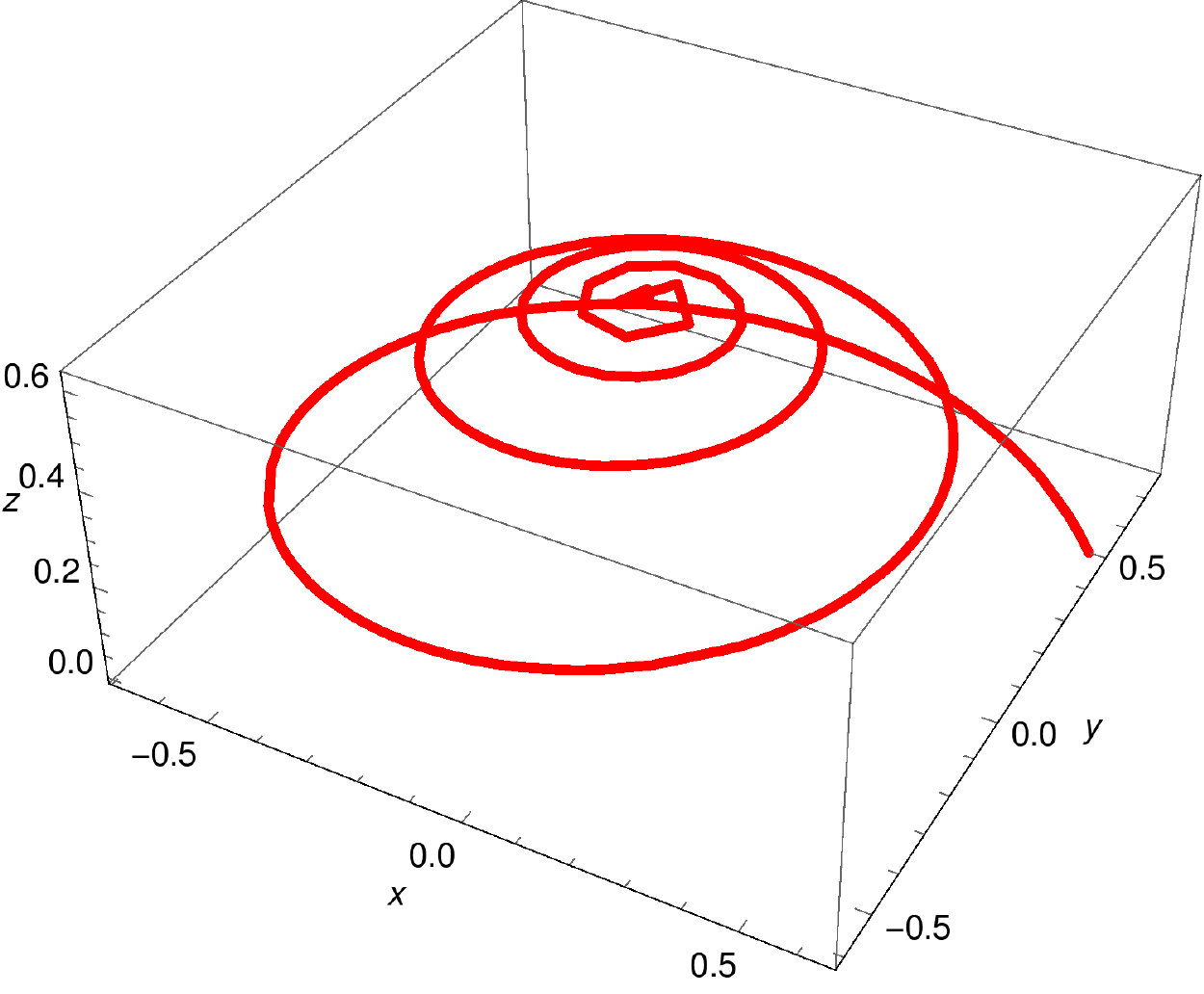} 
\includegraphics[scale=0.29]{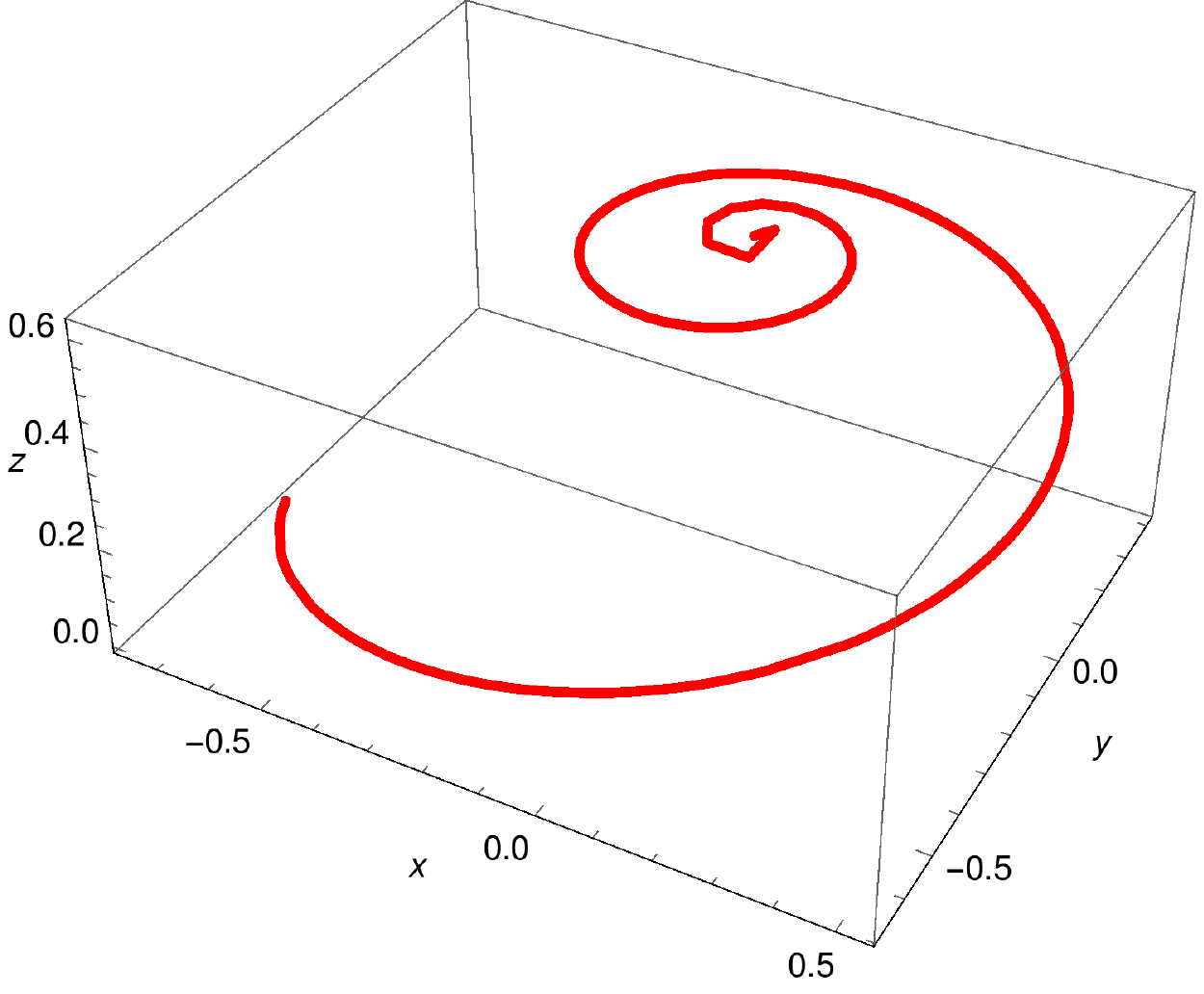} \\
(c)\includegraphics[scale=0.29]{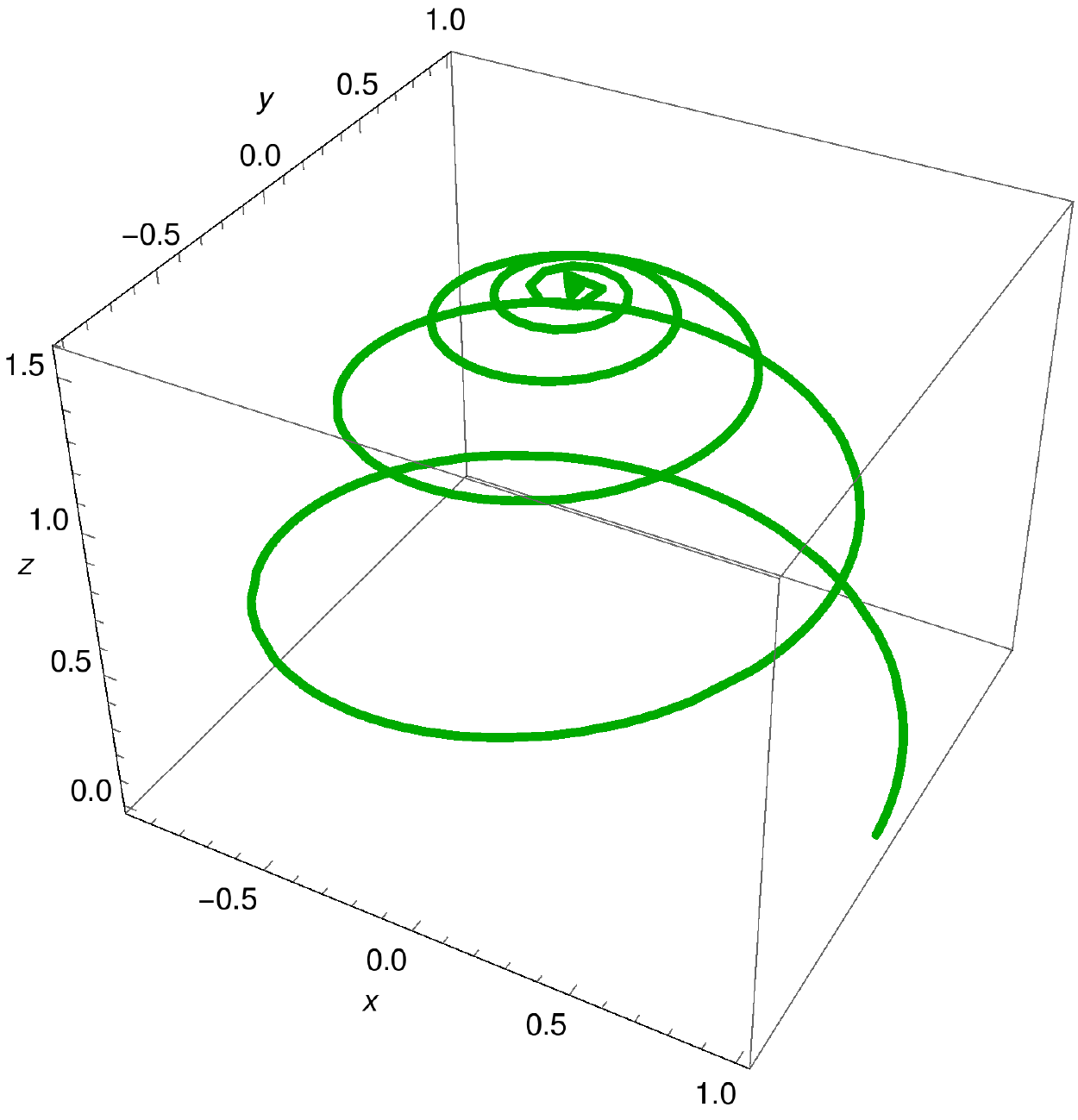}
\includegraphics[scale=0.29]{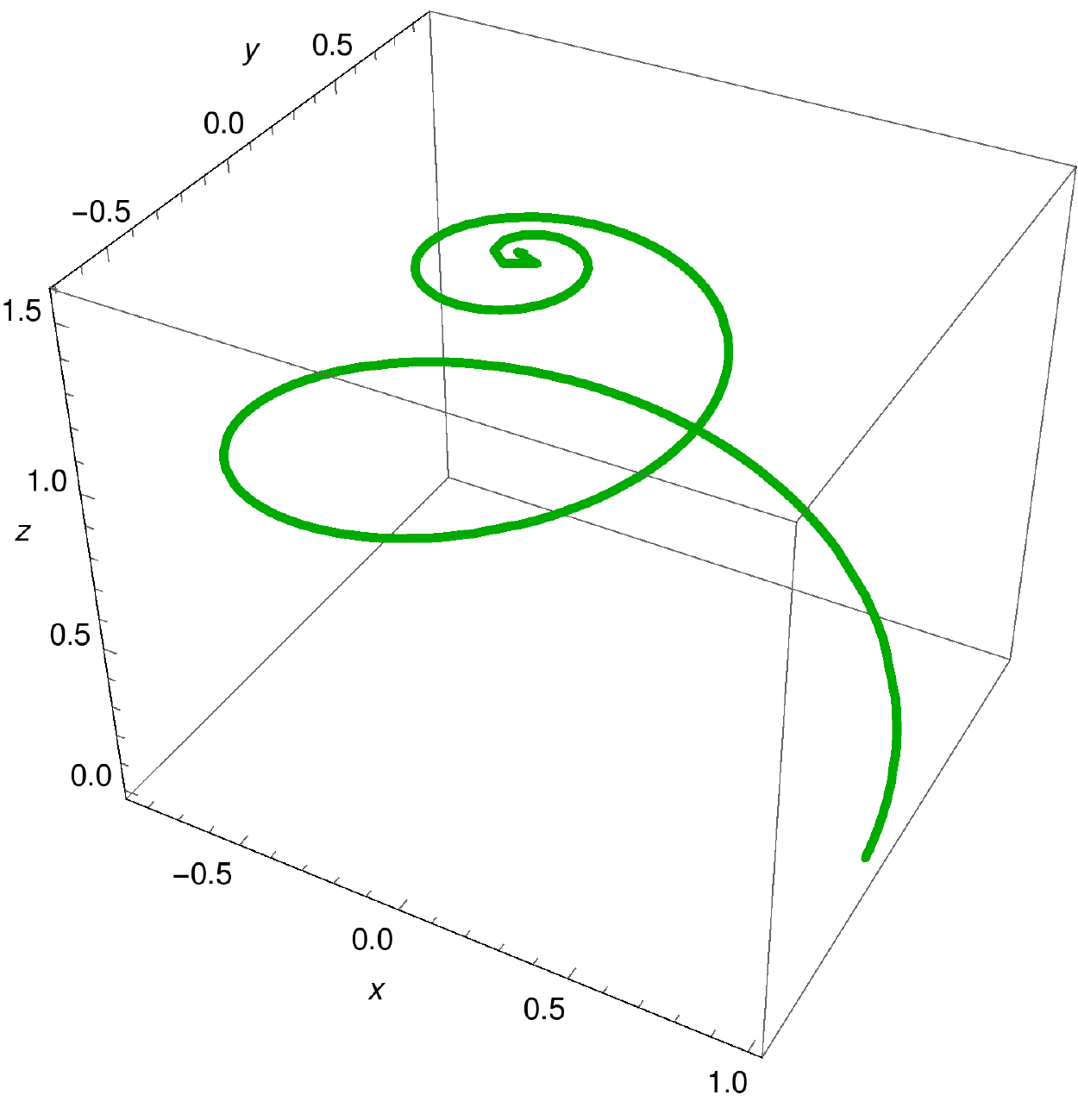}
\caption{Light trajectories in three dimensions for $\rho_M=1$,  $\rho_0=0.8$, $K=0.64$ and $\alpha=0.1,0.2$; the figures on the left  (right) side of  refer to $\alpha=0.1$ ($\alpha=0.2$). (a) Light path for rays of increasing radius starting at $\rho_0=0.8$. (b) Light path for rays of decreasing radius starting at $\rho_0=0.8$. (c) Light path for rays starting at $\rho_0=\rho_M$.}
\label{3d}
\end{figure}

The second family of solutions in (\ref{eqofparam}) is obtained when considering the negative sign, which corresponds to rays of decreasing radius: 
\begin{equation}
\rho(\zeta)=\sqrt{\rho_M^2-((\rho_M^2-\rho_0^2)^{1/2}+K\zeta)^2}. \label{solution-typeII}
\end{equation}
Here, the ray can decrease down to the defect core and hence be guided along the axis of the cylinder. When the light rays get closer to the defect core the angular momentum term in (\ref{energy-cons}) gets very large and hence, the light rays make more turns, see ({figs.} \ref{plotsr0} and  \ref{3d}).

By combining Eq. (\ref{solution-typeI}) with Eq. (\ref{pi2}), we obtain for the first family of solutions,
\begin{equation}
 {\phi}(\zeta)= \frac{1}{\alpha} {\tanh^{-1}\left(\frac{K\zeta-(\rho_M^2-\rho_0^2)^{1/2}}{\rho_M} \right)}\label{ang-typeI}
\end{equation}
and, by substituting Eq. (\ref{solution-typeII}) into Eq. (\ref{pi2}), we get  
\begin{equation}
 {\phi}(\zeta)= \frac{1}{\alpha}  {\tanh^{-1}\left(\frac{K\zeta+(\rho_M^2-\rho_0^2)^{1/2}}{\rho_M}\right)} \label{ang-typeII}
\end{equation}
for the second family of solutions.

By manipulating Eq. (\ref{solution-typeI}) and Eq. (\ref{ang-typeI}), we find $\rho=\rho(\phi)$  as being the confined Poinsot spiral:
\begin{equation}
\rho(\phi)=\frac{\rho_M}{\cosh{\alpha}{\phi}}. \label{eq-planar}
\end{equation}
By combining Eq. (\ref{solution-typeII}) and Eq. (\ref{ang-typeII}) we obtain the same Eq. (\ref{eq-planar}). With equation (\ref{eq-planar}) we can get the confined trajectories for light traveling in a plane $z=const$. We see that the smaller the value of $\alpha$, the stronger is the spiraling behavior as $1/\alpha$ can be understood as the ``spiraling strength'' (vorticity) of the defect \cite{fumeron2015optics,jacob2007semiclassical,figueiredo2017cosmology}. The same effect can be visualized in the three dimensional trajectories obtained from the parametric equations ${\rho}(\zeta)$ and ${\phi}(\zeta)$, see {fig.} \ref{3d}. For large values of $\alpha$, for example $\alpha=20$, the rays travel in nearly straight lines radially toward the defect core, see {fig. \ref{plotsr0}}.

We now examine the path followed by light in the HLCM device with a purely radial director field (see {fig.} \ref{fig1}(b)), which is obtained from homeotropic anchoring of the molecules at the boundaries. The metric tensor is given by:
\begin{eqnarray}
g_{ij}&=&\begin{pmatrix}
\alpha^2&0&0\\
0&-\rho^2&0\\
0&0&-1
\end{pmatrix} . \label{metric-b} 
\end{eqnarray}
The two Killing vectors are the same as in the circular case, $(\partial_{\phi})^i=(0,1,0)^{\rm T}$ and $(\partial_{\zeta})^i=(0,0,1)^{\rm T}$, associated with the covectors:
\begin{eqnarray}
&&(\partial_{\phi})_i=g_{ij}(\partial_{\phi})^j=(0,-\rho^2,0), \\
&&(\partial_{\zeta})_i=g_{ij}(\partial_{\zeta})^j=(0,0,-1).
\end{eqnarray}
The Killing equations are straightforwardly obtained as: 
\begin{eqnarray}
&&\zeta'=A, \\
&&\rho^2\phi'=C.
\end{eqnarray}
Energy conservation gives
\begin{equation}
\frac{\rho'^2}{2}-\frac{C^2}{2\alpha^2\rho^2}=\frac{B+A^2}{2\alpha^2}=E.  \label{energy-consII}
\end{equation}
The constant $E$ in Eq. (\ref{energy-consII}) is positive, then the solution is not exactly the same as Eq. (\ref{energy-cons}) where $E$ is negative. We can also see   that, since the first term in the left-hand side of Eq. (\ref{energy-consII}) cannot be zero, there is no turning point for the light trajectories (the trajectories are unbounded). Indeed, contrary to the first HLCM configuration {fig.} \ref{fig1}(a) light is not confined; if the ray starts with increasing $\rho$ at $\rho_0$ it will never decrease to the defect core, unless it is reflected by an outer layer. These arguments hold when looking at the parametric equation $\rho(\phi)\sim\frac{1}{\sinh{\alpha}{\phi}}$, which is another case of Poinsot's spiral, but  not of a confined type now. 

Hence, the first circular configuration should be favored to concentrate light. It must be highlighted that, in practice, such configuration also has the asset of preventing instabilities such as the escape in third dimension~\cite{CladisKleman72} that might break the director field and hence the guiding effects. 

\bigskip

To support the concentration effect, we are now examining the structure of the optical modes that propagate in the device (HLCM with circular director field, see {fig.} \ref{fig1}(a)). In the scalar wave approximation, the complex amplitude $\Phi$ of the wave is governed by the generalized form of the d'Alembert equation
\begin{equation}
\nabla_i\nabla^i\Phi-\frac{1}{c^2}\frac{\partial^2\Phi}{\partial t^2}=0,
\label{wave_eq_1}
\end{equation}
where $\nabla_i\nabla^i$ is the Laplace-Beltrami operator, whose action on the wave function $\Phi$ is given by
\begin{equation}
	\nabla_i\nabla^i\Phi=\frac{1}{\sqrt{|g|}}\partial_i\left(\sqrt{|g|}g^{ij}\partial_j\Phi\right).
    \label{Laplace-Beltrami-Op}
\end{equation}

In the case of harmonic time dependence of the form $\Phi(\rho,\phi,\zeta,t)=\psi(\rho,\phi,\zeta)\,e^{-i\omega t}$, Helmholtz equation is obtained from (\ref{Laplace-Beltrami-Op}) as
\begin{equation}
	-\frac{1}{\rho}\frac{\partial}{\partial \rho}\left(\rho\frac{\partial\psi}{\partial \rho}\right)+\frac{1}{\alpha^2 \rho^2}\frac{\partial^2 \psi}{\partial\phi^2}-\frac{\partial^2\psi}{\partial \zeta^2}+\frac{\omega^2}{c^2}\psi=0,
    \label{wave_eq_2}
\end{equation}
where $\omega$ is the angular frequency of the light. Using the ansatz $\psi(\rho,\phi,\zeta)=F_{\ell, k_\zeta}(\rho)e^{\pm i\ell\phi}
e^{\pm ik_\zeta \zeta}
$
 leads to the equation
\begin{equation}
	\rho^2\frac{d^2F_{\ell, k_\zeta}}{d\rho^2}+\rho\frac{dF_{\ell, k_\zeta}}{d\rho}-\left[\left(\frac{\omega^2}{c^2}+k_\zeta^2\right)\rho^2-\frac{\ell^2}{\alpha^2}\right]F_{\ell, k_\zeta}=0.
    \label{edo_r}
\end{equation}

Eq. (\ref{edo_r}) is the modified Bessel differential equation of imaginary order $i\ell/\alpha$, with solutions \cite{dunster1990bessel,olver2010nist}
\begin{equation}
	F_{\ell,k_\zeta}(\rho)=e_\ell\tilde{I}_{\ell/\alpha}\left(\bar{\omega}\rho\right)+f_\ell\tilde{K}_{\ell/\alpha}\left(\bar{\omega}\rho\right),
    \label{edo_r_sol}
\end{equation}
where $\bar{\omega}=\sqrt{k_\zeta^2+\omega^2/c^2}$ and $e_\ell,f_\ell$ are constants of integration. The functions $\tilde{I}_{\ell/\alpha}=\text{Re}I_{i\ell/\alpha}$ and $\tilde{K}_{\ell/\alpha}=K_{i\ell/\alpha}$ are linearly independent solutions of Eq. (\ref{edo_r}), with $I_{i\ell/\alpha},K_{i\ell/\alpha}$ being the modified Bessel functions of first and second kind, respectively \cite{olver2010nist}.
Note that, in an ordinary defect-free Euclidean space, the radial solution of Helmholtz equation would be a linear combination of  $ J_\ell(\rho\sqrt{\omega^2/c^2-k_\zeta^2})$ and $Y_\ell(\rho\sqrt{\omega^2/c^2-k_\zeta^2})$ instead  of $\Re ( I_{i\ell/\alpha}(\rho\sqrt{\omega^2/c^2+k_\zeta^2}))$ and $ K_{i\ell/\alpha}(\rho\sqrt{\omega^2/c^2+k_\zeta^2})$, which is the case here. We can see the effect in the solution of the metamaterial character of the metric via: i) the introduction of imaginary order Bessel  functions and ii) the change of Bessel to modified Bessel functions and the role of the defect amplitude through the appearance of $\alpha$ in the imaginary order of the modified Bessel functions.

\begin{figure}[h]
(a)\includegraphics[scale=0.48]{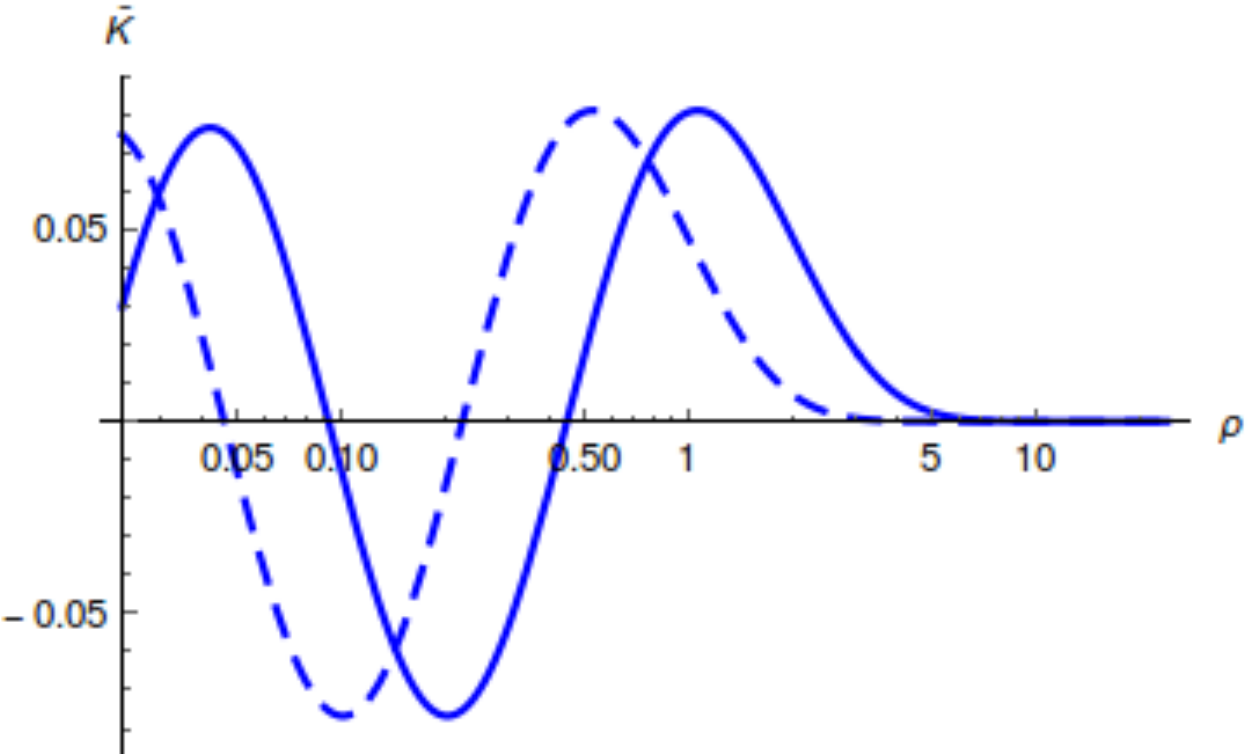} 
(b)\includegraphics[scale=0.48]{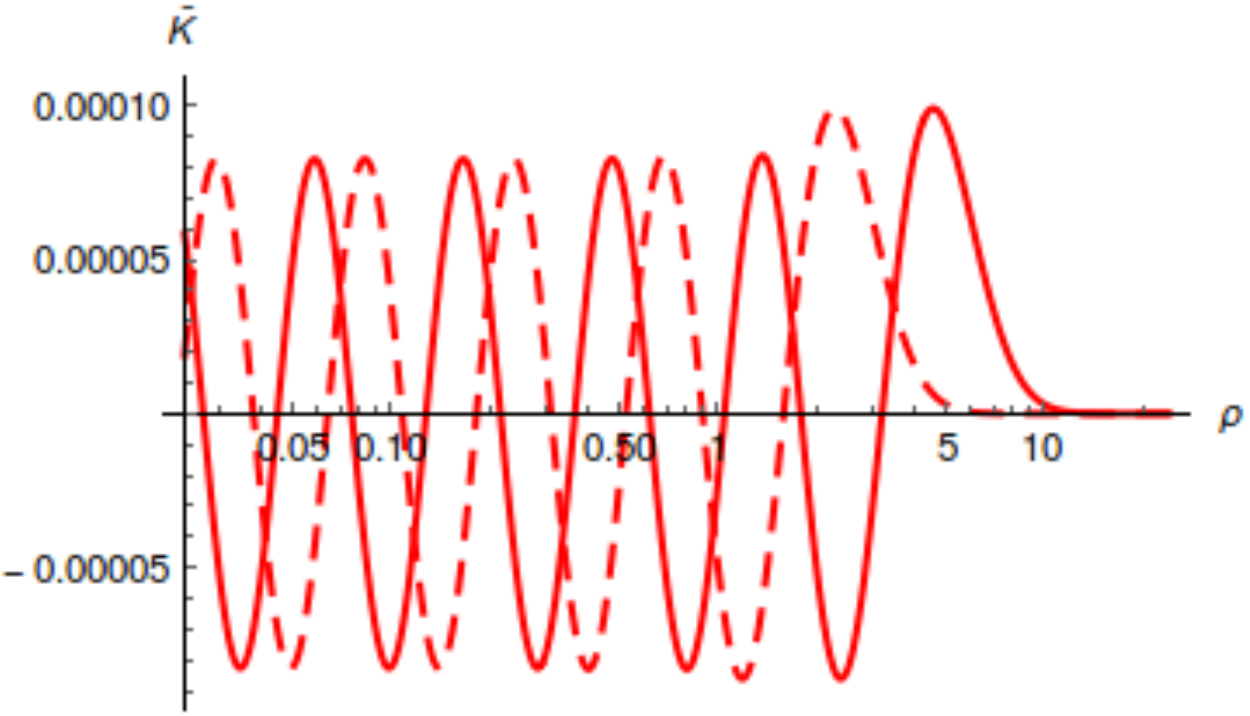} 
 \centering
\caption{The corresponding radial wave amplitudes $\tilde{K}$ for fixed $\ell=1$, $\alpha=0.17,0.5$ and $\bar{\omega}=1,2$. (a) for $\alpha=0.5$ and (b) for $\alpha=0.17$. The solid and dashed lines refers to $\bar{\omega}=1$ and $\bar{\omega}=2$, respectively. 
The smaller the value of $\alpha$, the more the fields oscillate near the origin. At large distances (logarithmic scale on the $x-$axis here) the behavior is exponential and it does not depend on $\ell/\alpha$.}
\label{i-bessel}
\end{figure}

The modified Bessel functions $\tilde{I}_{\ell/\alpha},\tilde{K}_{\ell/\alpha}$ oscillate rapidly near the origin, as one can see from their behaviors as $\rho\rightarrow 0^+$ \cite{dunster1990bessel,olver2010nist},
\begin{align}
\tilde{I}_{\ell/\alpha}(\bar{\omega} \rho)=&\left(\frac{\sinh(\pi \ell/\alpha)}{\pi \ell/\alpha}\right)^{1/2}\cos\left[\frac{\ell}{\alpha}\ln\left(\frac{\bar{\omega} \rho}{2}\right)-\gamma_{\ell/\alpha}\right]\nonumber\\
    &+O(\bar{\omega}^2\rho^2),\label{BesselI_near}\\
    \tilde{K}_{\ell/\alpha}(\bar{\omega} \rho)=&-\left({\frac{\pi\alpha/\ell}{\sinh(\pi \ell/\alpha) }}\right)^{1/2}\sin\left[\frac{\ell}{\alpha}\ln\left(\frac{\bar{\omega} \rho}{2}\right)-\gamma_{\ell/\alpha}\right]\nonumber \\
     &+O(\bar{\omega}^2\rho^2),\label{BesselK_near}
\end{align}
where $\gamma_{\ell/\alpha}$ is a constant defined as $\gamma_{\ell/\alpha}\equiv\arg\,\Gamma(1+i\ell/\alpha)$, with $\Gamma$ being the Gamma function. The rapid oscillations are due to the logarithmic argument of the trigonometric functions. Furthermore, for a fixed $\ell$, the smaller the value of $\alpha$, the stronger the oscillations become (reducing the value of $\alpha$ shrinks the period of the trigonometric functions). This behavior can be visualized in {figs. \ref{i-bessel}} (a) and (b).

Despite the similarities between $\tilde{I}_{\ell/\alpha}$ and $\tilde{K}_{\ell/\alpha}$ near the origin, their asymptotic behavior is exponential \cite{dunster1990bessel,olver2010nist},
\begin{align}
	\tilde{I}_{\ell/\alpha}(\bar{\omega} \rho)&=\left(\frac{1}{2\pi\bar{\omega} \rho}\right)^{1/2}e^{\bar{\omega} \rho}\left[1+O\left(\frac{1}{\bar{\omega} \rho}\right)\right],\label{BesselI_infty}\\
    \tilde{K}_{\ell/\alpha}(\bar{\omega} \rho)&=\left(\frac{\pi}{2\bar{\omega} \rho}\right)^{1/2}e^{-\bar{\omega} \rho}\left[1+O\left(\frac{1}{\bar{\omega} \rho}\right)\right].\label{BesselK_infty}    \end{align}
Hence, the first solution can be dismissed as it diverges at larges distances from the axis, which means that $e_\ell=0$. As expected from the geometrical optics limits, the field concentrates along the axis of the device.

\begin{figure}[h]
\vspace{3mm}
\centering
\includegraphics[scale=0.32]{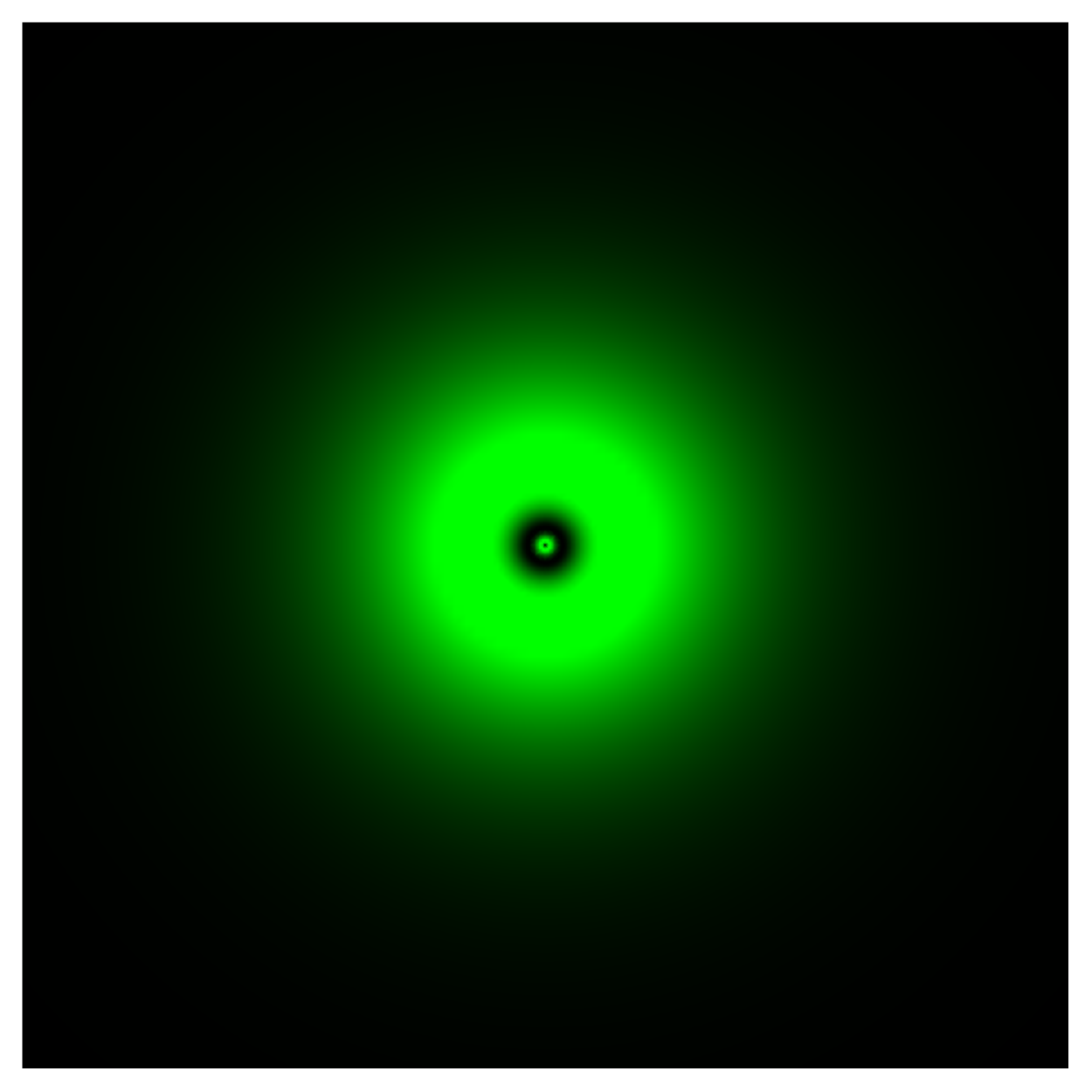}
\includegraphics[scale=0.32]{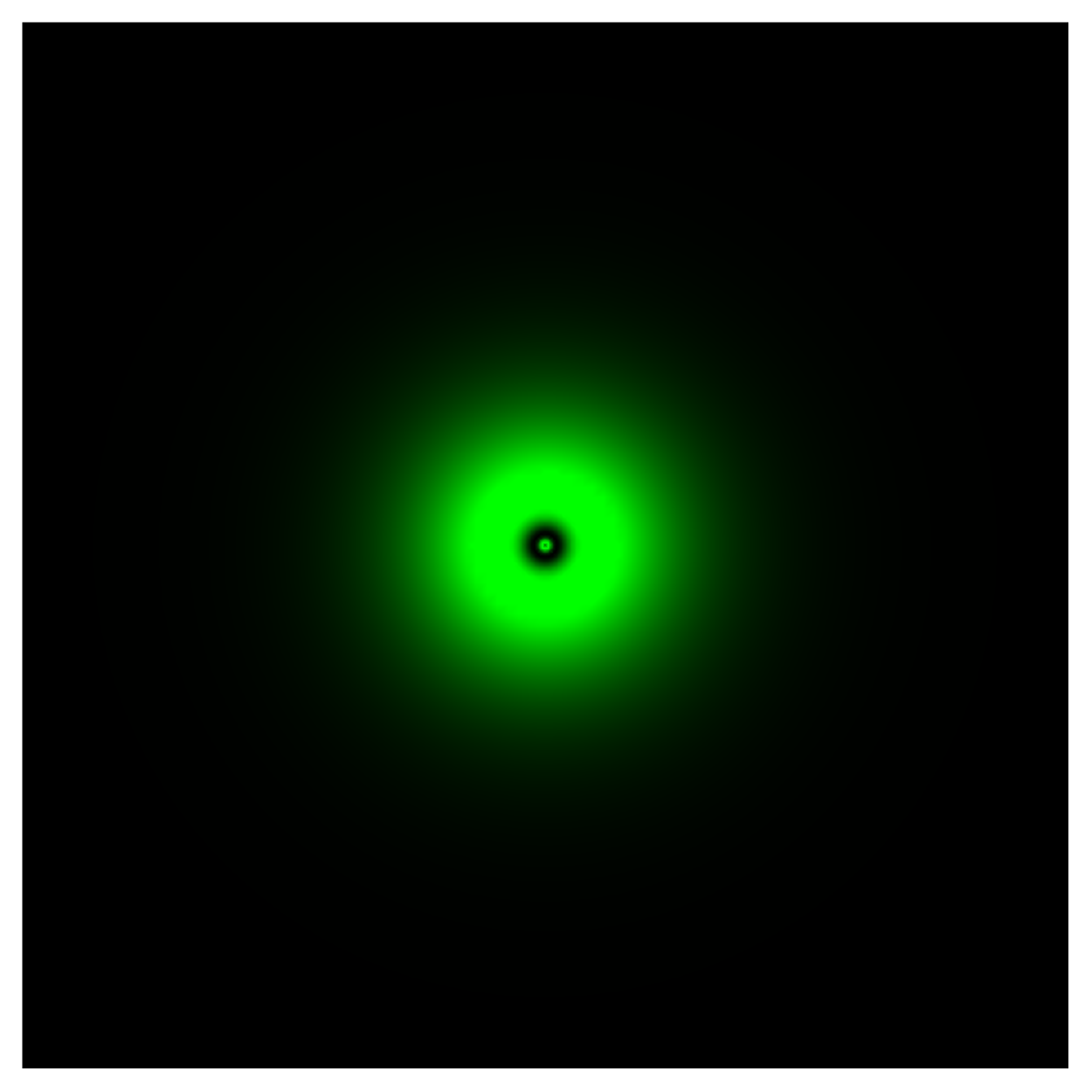}
\\
\includegraphics[scale=0.32]{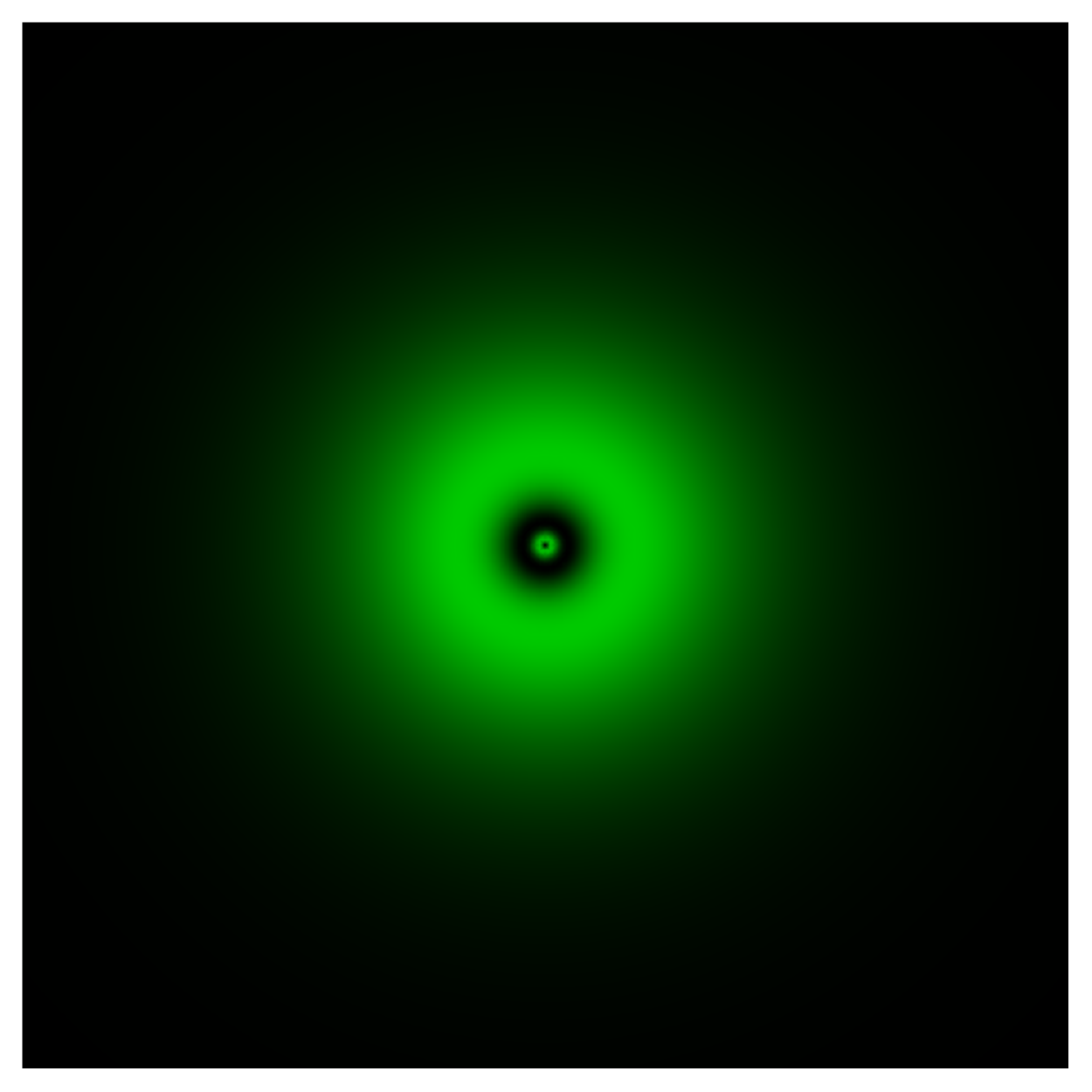}
\includegraphics[scale=0.32]{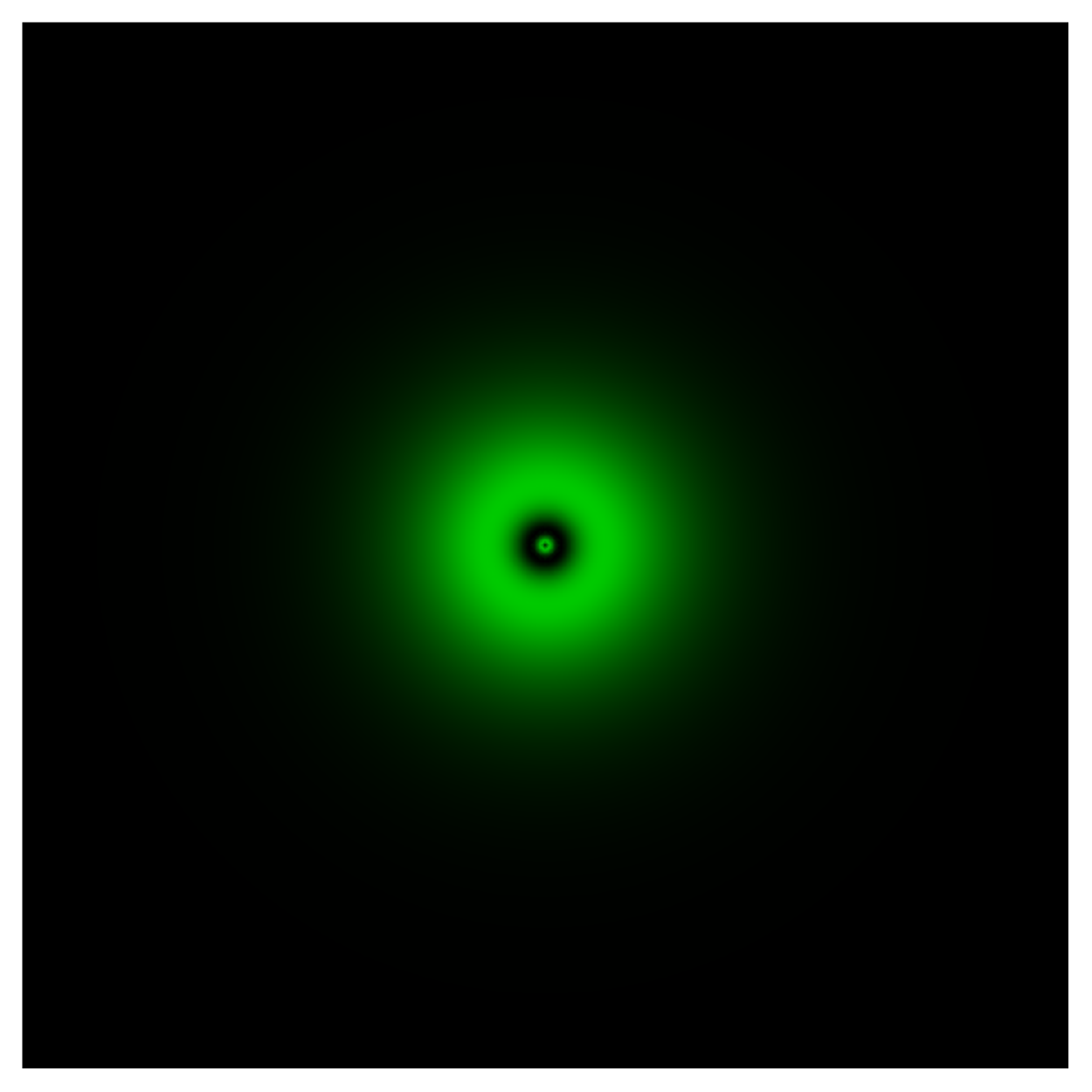}
\caption{$|\tilde{K}|^2$ intensity profiles, representing the transverse field distributions. All plots are in  the same scale and with $\ell=1$. The ones on the left (right) correspond to  $\bar{\omega}=0.67$ ($\bar{\omega}=1$). The top (bottom) ones correspond to $\alpha=0.90$ ($\alpha=0.83$).   
}
\label{intensity}
\end{figure}

For a given $z=const.$ plane, the intensity distribution for the propagating fields may be represented in terms of $|\tilde{K}_{\ell/\alpha}(\bar{\omega} \rho)|^2$, see {fig. \ref{intensity}}. We see that the bigger the value of $\alpha$ for a fixed $\ell$, the brighter the rings are (fields of high amplitudes). Besides, the bigger the value of $\bar{\omega}$, the smaller the light rings are (the more concentrated fields are the ones near the origin).  

\bigskip

We want to emphasize how powerful is the HLCM (circular  field configuration) for designing  optical concentrators. Both geometrical optics and wave optics treatments show that light is indeed concentrated while propagating in the HLCM medium. No matter how the rays are injected, they will fall down onto the defect core with or without being reflected in the outer layer of the cylindrical concentrator. In practice, it is shown that a given combination of materials forming metamaterials will keep hyperbolic properties only within a range of frequency defined in terms of the plasma frequency of the constituent materials \cite{smolyaninov2011critical,ferrari2015hyperbolic}.

{Finally, we remark that in this work we considered a real, albeit negative, refractive index.  In fact, in order to have a negative index, the permittivity must be complex, which is   provided by the metamaterial's metallic components. This implies a complex refractive index, whose imaginary part we  assumed to be small. The  main effect of the imaginary part of the refractive index in the device proposed here, is to damp the propagating beam, therefore decreasing its efficiency as a power transmitter but not impairing its concentrator feature.  Improvement of metamaterials in order to minimize such undesirable effects as ohmic losses, reflection and dispersion are presently the subject of very active research worldwide. Use of low-loss components, geometric tailoring, incorporation of gain media and directional scattering are some of the strategies pursued. }

\acknowledgments
F.M. is grateful to U. Lorraine, FACEPE, CNPq and CAPES for financial support. F.A. thanks the Coll\`ege Doctoral ``$\mathbb L^4$ collaboration'' (Leipzig, Lorraine, Lviv, Coventry) for financial support.

\bibliographystyle{ieeetr}
\bibliography{references.bib}

\end{document}